\documentclass[aps,prd,twocolumn,floatfix,preprintnumbers,nofootinbib,superscriptaddress]{revtex4-1}
\usepackage{mathtools}
\usepackage{graphicx}
\usepackage{color}
\usepackage{natbib}
\usepackage{multirow}
\usepackage{amsmath}
\usepackage[amssymb]{SIunits}
\usepackage{hyperref}

\newcommand{\fpbh}{f_{\rm PBH}}
\newcommand{\Mpbh}{M_{\rm PBH}}

\begin{document}
\preprint{LAPTH-023/22}
\title{The QCD phase transition behind a PBH origin of LIGO/Virgo events?}
\author{Joaquim Iguaz}
\email{iguaz@lapth.cnrs.fr}
\author{Pasquale D. Serpico}
\email{serpico@lapth.cnrs.fr}
\affiliation{LAPTh, CNRS, Univ. Savoie Mont Blanc, F-74940 Annecy, France}
\author{Guillermo Franco Abell{\'a}n}
\email{guillermo.franco-abellan@umontpellier.fr}
\affiliation{Laboratoire Univers \& Particules de Montpellier (LUPM), CNRS \& Universit{\'e} de Montpellier (UMR-5299)}
\date{\today}
\begin{abstract} 
The best-motivated scenario for a sizable primordial black hole (PBH) contribution to the LIGO/Virgo binary black hole mergers invokes the QCD phase transition, which naturally enhances the probability to form PBH with masses of stellar scale. We reconsider the expected mass function associated not only to the QCD phase transition proper, but also the following particle antiparticle annihilation processes, and analyse the constraints on this scenario from a number of observations: The specific pattern in cosmic microwave background (CMB) anisotropies induced by accretion onto PBHs, CMB spectral distortions, gravitational wave searches, and direct counts of supermassive black holes (SMBHs) at high redshift. We find that the scenario is not viable, unless an ad hoc mass evolution for the PBH mass function and a cutoff in power-spectrum very close to the QCD scale are introduced by hand. Despite these negative results, we note that a future detection of coalescing binaries involving  sub-solar PBHs has the potential to check the cosmological origin of SMBHs at the $e^\pm$ annihilation epoch, if indeed the PBH mass function is shaped by the changes to the equation of state driven by the thermal history of the universe.
\end{abstract}

\maketitle
\section{Introduction}
It is not unusual that the opening of a new astronomical window raises interesting questions for fundamental physics and/or cosmology. This is the case with the birth of gravitational wave astronomy, whose first ``heavy'' black hole merger event~\cite{LIGOScientific:2016aoc} has triggered a reflection on primordial black holes  (PBHs) being responsible for the bulk of these events~(see e.g.~\cite{Sasaki:2016jop,Bird:2016dcv,Ali-Haimoud:2017rtz,Kavanagh:2018ggo}).
PBHs, first proposed by Zeldovich $\&$ Novikov~\cite{novikov} and Hawking~\cite{hawking1}, could have formed in the early universe due to the collapse of large overdensities, although other formation mechanisms exist (see, for instance~\cite{carr1} and references therein).  They lead to a very rich phenomenology due to the wide range spanned in the associated parameter space.

A population of Schwarzschild PBHs is characterised by its mass function, in turn typically parameterised by its {\it shape} (i.e. relative abundances of  PBHs of different masses; in the simplest case a monochromatic distribution at a mass $\Mpbh$) and {\it normalisation}, often expressed as the fraction of DM in the form of PBHs $\fpbh\equiv \Omega_{\rm PBH}/\Omega_{\rm DM}$, with $\Omega_i$ denoting the cosmological average density of species $i$ in unit of the critical one.
A scenario with PBHs in the stellar mass range is especially interesting in the light of the LIGO/Virgo measurements of coalescing black hole binaries~\cite{LIGOScientific:2021djp}. Even if early proposals linking heavy BH merger events to PBHs constituting the totality of DM (see e.g.~\cite{Bird:2016dcv}) are not viable in the light of the reassessed constraints~\cite{carr1}, if PBHs with masses around $M_{\rm PBH} \sim {\cal O} (10) M_{\odot}$ contribute a fraction $\fpbh\simeq{\cal O}(10^{-3})$ to the DM of the universe, PBHs could explain a significant fraction of the events, improving the fits to the inferred mass distribution with respect to the simplest astrophysical sources templates, as discussed for instance in~\cite{Franciolini:2021tla}. 

PBH production models are hardly predictive on $\fpbh$, which is exponentially sensitive to the parameters. The PBH abundance is basically used as a free fitting coefficient in quantitative phenomenological studies. A similar parametric approach might be followed of course for the shape of the mass function required to fit the data, at the expense of the model falsiability.
One may wonder, however, if this shape could be theoretically motivated, since this aspect would be amenable to observational tests. Interestingly, for the mentioned stellar mass scale, i.e. $\Mpbh \sim 0.1\,M_{\odot}-100\,M_{\odot}$, this is indeed the case if the change in thermodynamical properties associated to the  quantum chromodynamics (QCD) phase transition in the early universe is taken into account. Proposals in this sense abound, see e.g.~\cite{JEDAMZIK1998155,PhysRevD.59.124014,10.1093/mnras/stw2138,2018,Carr:2019hud,Carr:2019kxo,Jedamzik:2020omx}, with the earliest ones even pre-dating the discovery of gravitational wave events. 
 
 In this work, we revisit this ``best motivated'' scenario to assess its viability in the light of current constraints from  cosmic microwave background (CMB) anisotropies associated to accretion onto PBH~\cite{accretion}, from CMB spectral distortions~\cite{Chluba_2012}, as well as null searches of sub-solar PBHs ~\cite{nitz2022broad} and a stochastic gravitational wave background~\cite{KAGRA:2021kbb} in LIGO/Virgo. To do so, we compute the expected mass function associated not only to the QCD phase transition proper, but also the following particle antiparticle annihilation processes, down to the electron-positron annihilation  taking place later in the cosmic history of the universe. This implies a peculiar mass function with features extending up to $M_{\rm PBH} \sim 10^{7}\,M_{\odot}$. 
 
The paper is organized as follows: In Section~\ref{sectionII} we  recap the early universe physics relevant for the PBH mass distribution, and (re)derive the key relations. In Section~\ref{sectionIII} we discuss their implications on the PBH abundance, present our results, and assess the viability of the scenario under study. In Section~\ref{sectionIV}, we  discuss possible loopholes and conclude.
Bounds from CMB spectral distortions are separately treated in Appendix~\ref{appA}. This is because  it is typically argued in modern literature that the bounds due to CMB spectral distortions can be lifted (naively, arbitrarily much) by invoking larger and larger non-gaussianities (NGs), see for instance~\cite{PhysRevD.97.043525}. In  Appendix~\ref{appA}, we revisit the physics behind this claim and find that for extremely large NGs the argument should break down, with the exclusion bound eventually stronger than previously suggested. Finally, in Appendix~\ref{appB}, we derive CMB constraints on disk-accreting PBH including extended mass functions, in order to gauge the error that is introduced when simply recasting the existing constraints for monochromatic functions with a linear approximation.

\section{Preliminaries}\label{sectionII}
How easily PBHs can form via gravitational collapse from a given power spectrum of fluctuations depends on the equation of state $w\equiv P/\rho$ in the early universe, which controls how well pressure can oppose gravity, as reviewed for instance in~\cite{2018}.  All other conditions being the same, any drop in $w$ is thus expected to map directly into an enhancement of the PBH production. In sec.~\ref{sectionIIA} we review the evolution of $w$ vs. temperature $T$, while in Sec.~\ref{sectionIIB} we focus on how to translate $w(T)$ into the PBH mass function.

\subsection{Equation of state in the early universe}\label{sectionIIA}

\begin{figure}[htbp]
  \centering
    \includegraphics[width=0.45\textwidth]{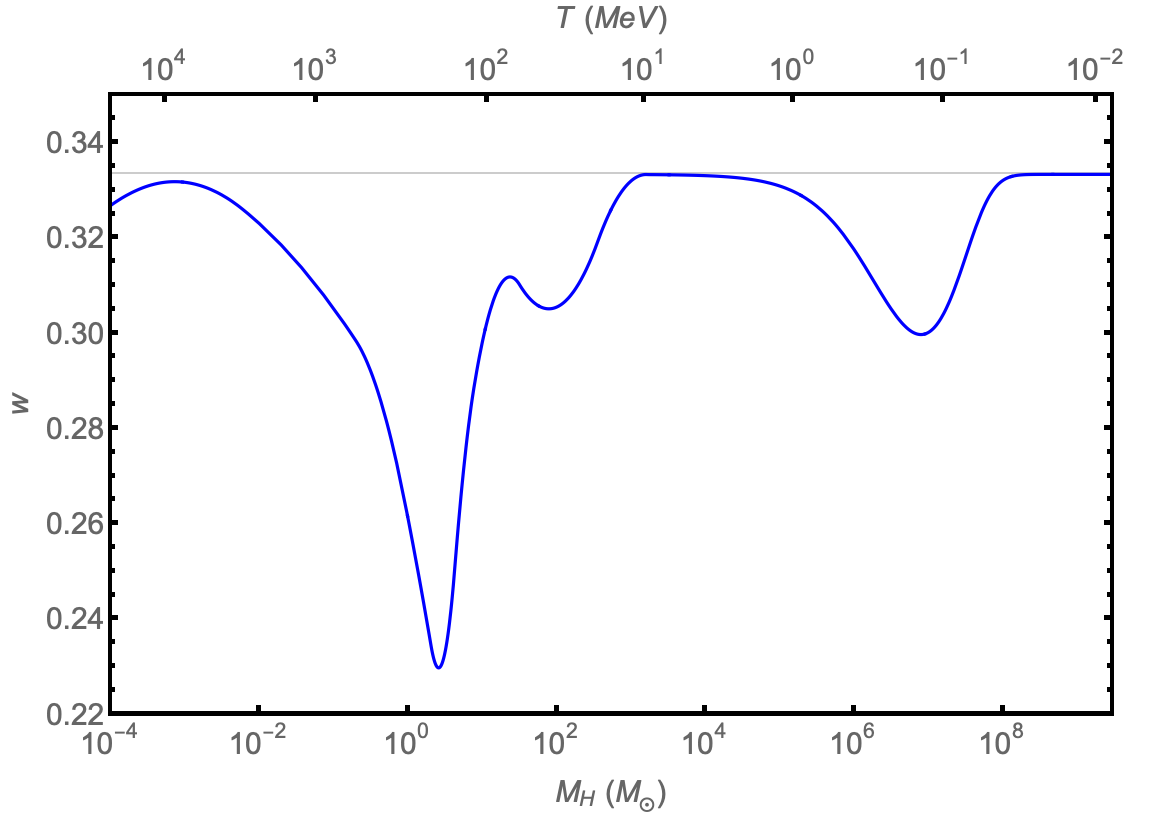}
    \caption{Equation of state parameter $w$ as a function of the temperature of the universe (top scale) or Hubble mass $M_{H}$ (bottom scale). The gray horizontal line corresponds to the value during radiation domination $w=1/3$.}
    \label{fig1}
\end{figure}

The standard value $w=1/3$ during the radiation domination phase changes due to the evolution of the effective number of relativistic degrees of freedom. These are defined in terms of energy density $\rho$, entropy density $s$, and temperature $T$ as~\cite{KT}
\begin{equation}
\begin{split}
    g_{\rm eff}(T) & \equiv\frac{30 \rho}{\pi^{2}T^{4}}\,, \\ h_{\rm  eff}(T) & \equiv\frac{45 s}{2 \pi^{2}T^{3}}\,,
    \label{eq1}
\end{split}
\end{equation}
and decrease as the Universe cools down. This is a consequence of particles annihilating out of the plasma when the temperature of the Universe decreases below the corresponding mass thresholds. This effect induces dips in $w(T)$, as we report in Fig.~\ref{fig1}.
In both cases, we also report the functions of interest with respect to another scale, the mass  $M_H$ enclosed in the Hubble horizon, introduced in the following (see Eq.~\eqref{eq5}) and which is a close proxy of the PBH mass scale that can be produced at that epoch. From the relation $P=sT-\rho$ and using Eq.~\eqref{eq1}, one can easily obtain
\begin{equation}
    w(T)=\frac{4 h_{\rm eff}(T)}{3 g_{\rm eff}(T)}-1\,.
    \label{eq2}
\end{equation}
In detail, for $g_{\rm eff}$ and $h_{\rm eff}$ we use the values in Table S2 of~\cite{Borsanyi:2016ksw}  and Table A1 of~\cite{Husdal:2016haj} to derive the equation of state for the Standard Model d.o.f.'s via Eq.~\eqref{eq1}.

Note that the first dip at $M_H\sim 0.01\div 10\, M_{\odot}$ is caused by the QCD phase transition proper, with the confinement of quarks and gluons into hadrons responsible for a  large drop in the number of the dof's at around $T_{\rm QCD}\sim 150$ MeV. 
The second dip around $M_H\sim 100 M_{\odot}$ is associated to the disappearence of the pion and muon dof's from the plasma. The third dip at $M_H\sim 10^{7} M_{\odot}$ is due to the electron-positron annihilation, the last such episode in the standard thermal history of the universe.

\subsection{PBH mass distribution}\label{sectionIIB}

Qualitatively, there will be an enhanced probability to form PBHs from  the collapse of horizon patches enclosing a mass $M_{H}$, coinciding with each drop of $w(M_{H})$ in Figure~\ref{fig1}. This translates into a drop of the critical threshold needed for collapse, $\delta_{c}$. Specifically, we map the drops in $w$ into drops in $\delta_c$ using the results of Figure 8 in~\cite{2013}. 

\begin{figure}[htbp!]
    \centering
    \includegraphics[width=0.45\textwidth]{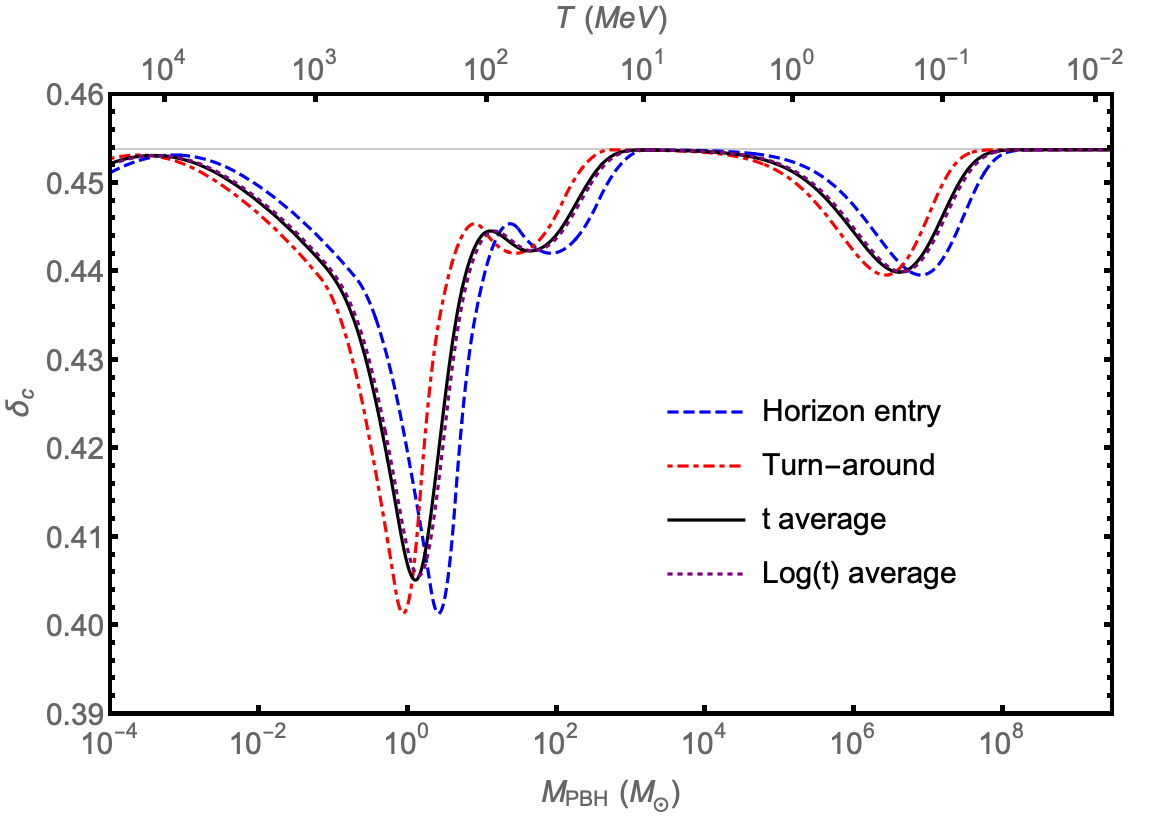}
    \caption{Critical density perturbation for collapse $\delta_{c}$ as a function of the $M_{H}$ for the four different prescriptions mentioned in the text. The gray horizontal line corresponds to the reference value of $\delta_{c}$ for pure radiation, $w=1/3$.}
    \label{fig2}
\end{figure}

There are slightly different options for the PBH criterion formation~\footnote{In the main text we stick to linear relations and the simplified picture of PBH formation, since it is sufficient to develop our argument. Further refinements are discussed in Appendix~\ref{appA}.}. Physically, even if a critical overdensity corresponding to a mass $M_{H}$ is attained when a given mode of curvature perturbation reenters the horizon, a finite  time elapses until the  PBH actually forms. In the meanwhile, the horizon mass grows. Following the discussion in~\cite{2018}, in order to bracket this uncertainty in the relation $\Mpbh$-$\delta_{c}$, in Figure~\ref{fig2} we consider four different prescriptions for the computation of $\delta_{c}$: at horizon entry, at turn-around time, time averaged value, and logarithmic time averaged value.

Note that the time averaged profile is very close to the logarithmic time averaged one, and is also in between the other cases; henceforth, unless stated  otherwise, we will take the results obtained from the time averaged profile as our benchmark results, keeping in mind that different prescriptions can lead to differences within a factor 2 of the reference value. Using an alternative benchmark, such as that PBH mass only amounts to 70\% of the horizon mass, as in~\cite{Carr:2019kxo}, is within this uncertainty bracket. The slight shift to lower mass  and broadening of the mass function due to criticality in PBH formation~\cite{Kuhnel:2015vtw} is also of comparable size and of minor impact for the already broad mass function of interest here, see the discussion in~\cite{Carr:2019kxo}.

The threshold value $\delta_{c}$ can then be used to compute the fraction of the Universe collapsing into PBHs as
\begin{equation}
    \beta=2 \int_{\delta_{c}}^{\infty} {\rm d}\delta \frac{M}{M_{H}}P(\delta)\,,
    \label{eq3}
\end{equation}
where $P(\delta)$ is the probability density function of the density contrast, which is typically assumed to be Gaussian. How this is generalised to a non-Gaussian case is discussed in Appendix~\ref{appA}.
The variance of $P(\delta)$ is related to the power spectrum of energy density fluctuations $\mathcal{P}_{\delta}(k)$ via
\begin{equation}
    \sigma^{2} =\int_{0}^{\infty} W(kR)^{2}\mathcal{P}_{\delta}(k)\frac{{\rm d} k}{k}\,,
        \label{eq4}
\end{equation}
where
\begin{equation}
    \mathcal{P}_{\delta}(k) =\frac{16}{81} (kR)^{4} \mathcal{P}_{\zeta}(k) \,.       \label{eq4b}
\end{equation}
In Eq.~\eqref{eq4}, $W(kR)=\text{exp}(-\frac{(kR)^{2}}{4})$ is the Fourier transform of the window smoothing function ~\cite{2021}, $R=1/k^{*}$ is the size at the time the mode $k^*$ enters the horizon, which is related to $M_{H}$ via
\begin{equation}
    M_{H}=17 \left(\frac{g}{10.75}\right)^{-1/6} \left(\frac{k^{*}}{10^{6} \text{Mpc}^{-1}}\right)^{-2} M_{\odot}\,.
    \label{eq5}
\end{equation}
Eq.~\eqref{eq4b}, expressed in terms of the curvature power spectrum $\mathcal{P}_{\zeta}(k)$,  follows from the linear-order relation between the overdensity and curvature perturbation $\zeta_k$ in Fourier space
\begin{equation}
    \delta_{k}=\frac{2(1+w)}{(5+3w)} \left( \frac{k}{aH} \right)^{2} \zeta_{k},
    \label{eq6}
\end{equation}\\
once we replace  $w=1/3$. 
Given a Gaussian distribution for $P(\delta)$, $\beta$ can be written as
\begin{equation}
    \beta=\text{erfc}\left( \frac{\delta_{c}}{\sqrt{2 \sigma^{2}}} \right),
    \label{eq7}
\end{equation}
which allows us to compute the total fraction of DM in the form of PBHs as
\begin{equation}
\begin{split}
\fpbh= & \int \psi_p(M){\rm d} M\equiv\int F(M)\frac{{\rm d} M}{M}
    = \\
   & \int \left( \frac{M}{M_{\rm eq}} \right)^{-1/2} \frac{\beta(M)}{\Omega_{\rm DM}}\frac{{\rm d} M}{M},
    \label{eq8}
\end{split}
\end{equation}
where the label $p$ implicitly represents the additional parameters entering the underlying power spectrum. 
In Eq.~\eqref{eq8}, $M_{\rm eq}=2.8\times 10^{17} M_{\odot}$ is the mass within the horizon at matter-radiation equality, and we assumed that all the mass within the horizon is eventually ending up in the PBH. 

\subsection{Extended mass function formalism}\label{sectionIIC}

Upper bounds on $\fpbh$ vs. $M$ are typically obtained assuming a monochromatic mass function; let us denote this function $f_{\rm mono}^{\rm max}(M)$. However, as can be seen from Eq.~\eqref{eq8}, the scenario studied in this work naturally yields extended mass distributions for the PBHs. Therefore, we are interested in revisiting previous bounds in the literature that were derived under the assumption of monochromaticity. A quick recasting of the existing bounds for an extended mass function can be obtained under linear hypotheses according to the procedure described in~\cite{extmass}. The bounds on $\psi_p(M)$ and thus $\fpbh$ (see eq.~\eqref{eq8}) are obtained via the condition
\begin{equation}
    \int_{M_{\rm min}}^{M_{\rm max}}  {\rm d} M \frac{\psi_p(M)}{f_{\rm mono}^{\rm max}(M)}=1,
    \label{eq11}
\end{equation}
where $M_{\rm min}$ ($M_{\rm max}$) is taken as the minimum (maximum) value for which the monochromatic bound has support.

For comparison purposes, we find also useful to define the fraction of DM in the form of PBHs in the range yielding coalescence events needed to account for ``heavy'' mergers in LIGO/Virgo, defined as
\begin{equation}
    f_{\rm GW}\equiv \int_{5M_{\odot}}^{160M_{\odot}} \psi_p(M)  {\rm d}  M.
    \label{eq12}
\end{equation}
Obviously, one has $f_{\rm GW}\leq f_{\rm PBH}$.

In Appendix~\ref{appB}, we have compared the results of the linear estimate with dedicated numerical calculations, for the case of CMB anisotropy bounds from PBH accretion. This is a useful exercise since the very large support of the mass function entering especially CMB anisotropy bounds can cast doubts on the reliability of the linear approximation.

\section{Results}\label{sectionIII}

Even accounting for the change in the EOS experienced in the early universe, if the primordial power spectrum (PS) were as small as the one extrapolated to small spatial scales from the fits to the CMB, the predicted amount of PBHs would be negligible. Thus, proponents of the ``QCD-inspired'' scenario for stellar mass PBH require more or less explicitly that the PS is enhanced at wavenumbers $k>k_\bullet$ with $ k_{\rm QCD}\gg k_\bullet\gg k_{\rm CMB}$. Setting $k_\bullet \sim k_{\rm QCD}$ is of course equivalent to introduce by hand a feature that is responsible for the PBH production, and is essentially equivalent to deny a crucial role to the mechanism for PBH production that we are testing here. How to model the PS at $k>k_\bullet$ is unspecified in most scenarios, but an agnostic choice (and often found in the literature, see e.g.~\cite{2018}) is to consider that the PS is again close to scale invariant at these small scales. For illustration purposes, following Figure 5 in~\cite{2018}, in Figure~\ref{fig3} we plot the results for a variance of the PS scaling with mass as
\begin{equation}
    \sigma^{2}=0.0033 \left( \frac{M}{10 M_{\odot}} \right)^{n_{M}},
    \label{eq13}
\end{equation}
where $n_M=0.025$ ($n_M=0$ corresponds to the scale invariant limit). As long as $|n_M|$ is not too large (say, of the same order of $|n_s-1|$ setting the departure from scale-invariance at CMB scales) the following considerations are not crucially dependent from $n_M$. (If this is not the case, the rationale of the proposal we are discussing would be shattered, since the mass function would be heavily dependent of the peculiar physics introduced at these scales, of course.)
\begin{figure}[htbp!]
    \centering
    \includegraphics[width=0.45\textwidth]{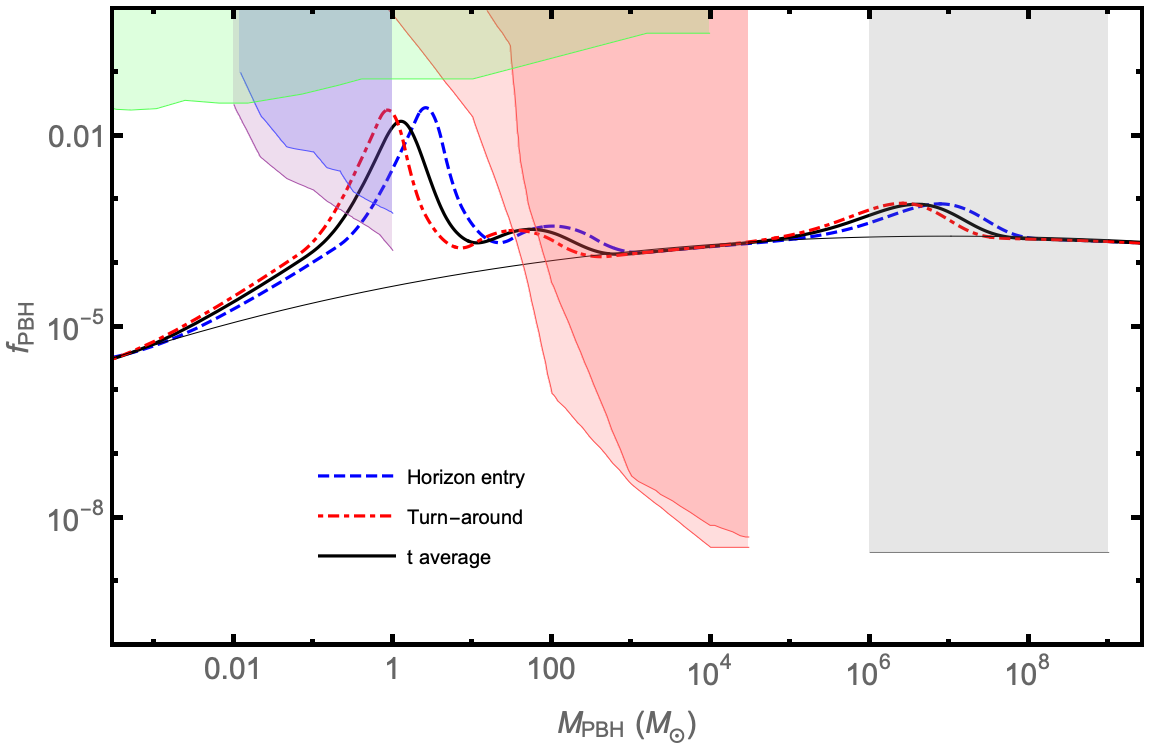}
    \caption{PBH mass distribution (corresponding to the function $F(M)$ in Eq.~\ref{eq8}) for a quasi-flat spectrum with a spectral index $n_{M}=0.025$. The thin black line corresponds to the scenario without QCD/$e^{+}e^{-}$ enhancement. It corresponds to Figure 5 in~\cite{2018}. We also plot excluded regions from microlensing~\cite{PhysRevD.99.083503}\cite{Tisserand_2007}\cite{Alcock_2001} in light green, GW production~\cite{nitz2022broad} for two different two-point delta mass distributions in blue and purple, accretion effects on CMB anisotropies~\cite{accretion} in pink/red and inferred SMBH population at high redshift~\cite{accretion} in gray.}
    \label{fig3}
\end{figure}
For the sake of illustration, the mass distributions displayed in Figure~\ref{fig3} are obtained by requiring that the main peak at $M\sim 1M_{\odot}$ reaches the benchmark value derived from GW data, roughly $f_{\rm GW}\sim 10^{-3}$~\cite{Franciolini:2021tla}. We also plot some other constraints from null gravitational waves searches, cosmology and astrophysics. 
Clearly, benchmarking to fits to coalescing BH rates from GW observations risks to lead to tensions with existing constraint. 
A parametric way to evade the constraints would consist in pushing $k_\bullet$ closer and closer to $k_{\rm QCD}$. This is the way we decide to gauge the credibility of the scenario in absence of fine-tuning.  In particular, we parameterize the enhanced PS at small scales as
\begin{equation}
\begin{split}
    \mathcal{P}_{\zeta}(k)&=\mathcal{P}_{\rm CMB}(k)+\frac{\Delta}{1+\text{exp}\left(\frac{k_{\bullet}-k}{\text{Mpc}^{-1}}\right)} \\
    \text{with}  \quad &\mathcal{P}_{\rm CMB}(k)=A_{s}\left(\frac{k}{0.05 \text{Mpc}^{-1}}\right)^{n_{s}-1},
    \label{eq14}
\end{split}
\end{equation}
with $k_\bullet$ and $\Delta$ two PS parameters that can be traded for the phenomenological more appealing $M_{\rm cut}=( \frac{k_{\bullet}}{10^{6} \text{Mpc}^{-1}} (\frac{g_{*}}{10.75})^{1/12} 17^{-1/2})^{-2} M_{\odot}$ (the mass scale above which the PBH mass function is cut by hand) and $\fpbh$, computed as described in the previous section. 
We fix the PS parameters at large spatial scales respectively to $A_{s}=2\cdot 10^{-9}$ and $n_{s}=0.965$ ~\cite{planck2020}.

\begin{figure}[htbp!]
    \centering
    \includegraphics[width=0.45\textwidth]{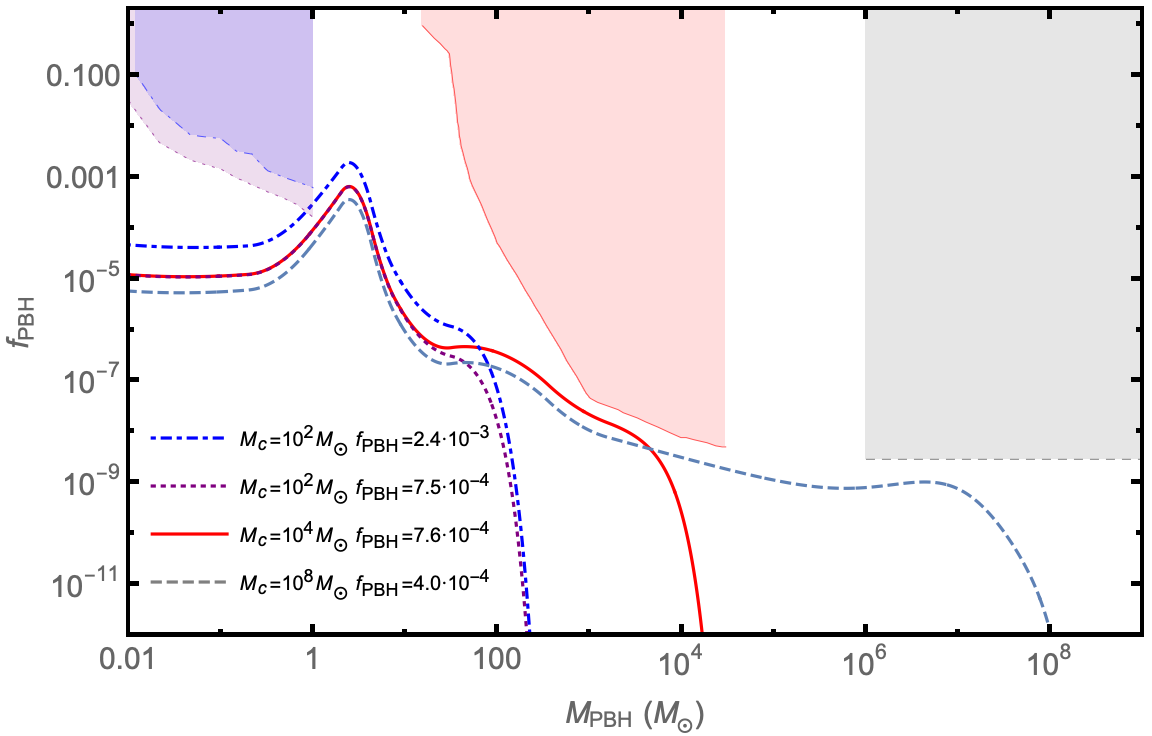}
    \caption{Mass functions (again, corresponding to the function $F(M)$ in Eq.~\ref{eq8}) consistent with three different sets of bounds for fixed values of $\text{M}_{c}$. We show the results for $\text{M}_{c}=10^{8} M_{\odot}$ and SMBH counting (gray), $\text{M}_{c}=10^{4} M_{\odot}$ and spherical accretion (red) and $\text{M}_{c}=10^{2} M_{\odot}$ and GW production (blue and purple). See Figure~\ref{fig3} for the corresponding references.}
    \label{fig4}
\end{figure}

In Figure~\ref{fig4}, we illustrate how, depending on the assumed value of $M_{\rm cut}$, the model is subject to a different set of constraints. If $M_{\rm cut}\gtrsim 10^6\,M_\odot$, it was argued in~\cite{Serpico:2020ehh} that the mass in PBHs would exceed the one inferred in supermassive black holes (SMBH) at $z\gtrsim 6$ from astrophysical observations when $\fpbh(M\geq 10^6\,M_\odot)>2.9\times 10^{-9}$. The origin of these objects is still unclear, and it is conceivable that they might be primordial, see e.g.~\cite{Inayoshi:2019fun} for a review. For the mass function predicted in the scenario considered here, and assuming $M_{\rm cut}= 10^8\,M_\odot$, this leads to the bound $\fpbh<4.0\times 10^{-4}$ (dashed gray curve in Fig.~\ref{fig4}). In this case, the model is also subject to tight constraints coming from CMB spectral distortion, as discussed in Appendix~\ref{appA}. 

One can get rid of this constraint if lowering  $M_{\rm cut}$. Still, the scenario is subject to  the bound from CMB anisotropies. For $M_{\rm cut}=10^4\,M_\odot$, for instance, we obtain $\fpbh<3.9\times 10^{-4}$ for the disk  accretion model considered in~\cite{accretion,Serpico:2020ehh}, or $\fpbh<7.5\times 10^{-4}$ for the spherical accretion model of~\cite{accretion,Serpico:2020ehh},
if we used the approximated linearised model to deal with the extended mass function. In Appendix~\ref{appB}  we have compared the results of the linear estimate with dedicated numerical calculations accounting for the realistic form of the extended mass function. These calculations indicate that the actual bounds are a factor $\sim 1.5$  stronger than the simpler  estimates, so that our quoted bounds are, if anything, on the conservative side. 

\begin{figure}[htbp!]
    \centering
    \includegraphics[width=0.45\textwidth]{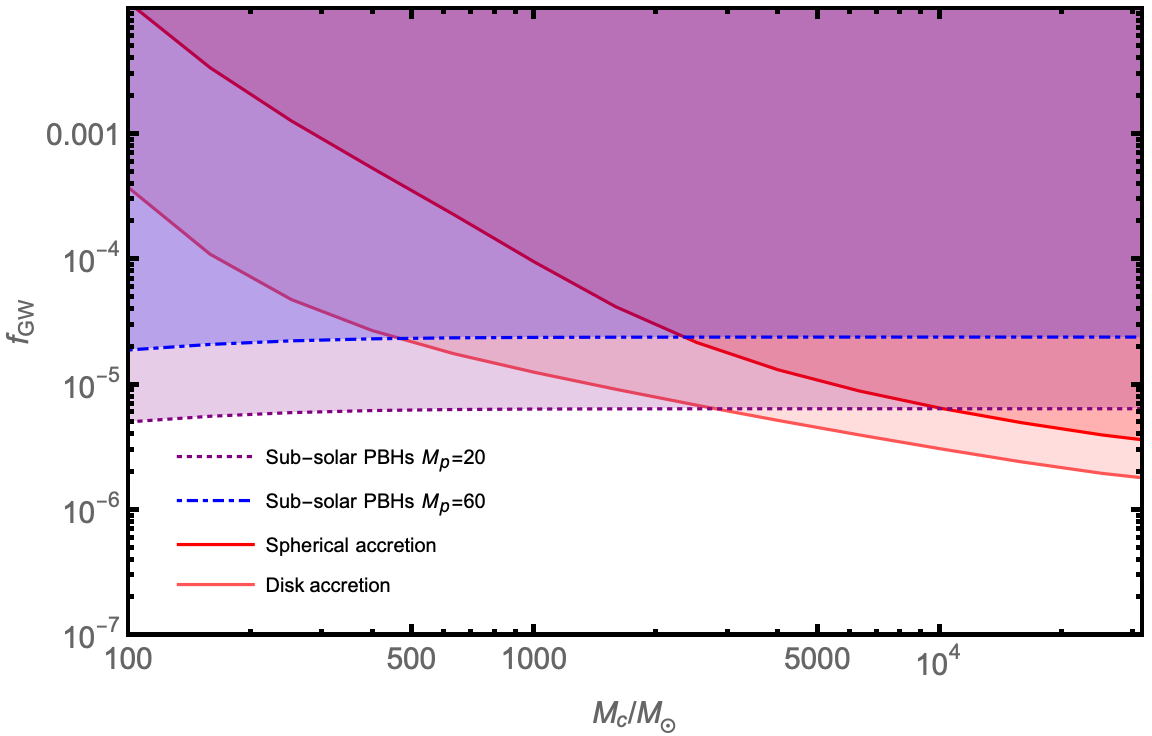}
    \caption{Upper bounds on $f_{\rm GW}$ vs the cutoff mass $M_{c}$ from CMB anisotropies (pink/red excluded regions) and non-observations of mergers with a BH whose mass is sub-solar; these bounds mildly depend on the heavier partner mass $M_{p}$, and the two blue bands in the plot  bracket the extremes; see \cite{nitz2022broad} for more details.}
    \label{fig5}
\end{figure}

Further lowering $M_{\rm cut}$ relaxes the CMB anisotropy bounds, significantly so if $M_{\rm cut}\lesssim {\cal O}(10^2)\,M_\odot$ (see Fig.~\ref{fig5}), i.e. when we cut the mass function just above the heaviest BH detected by LIGO/Virgo, i.e. at a scale $k_\bullet$  within an order of magnitude of the QCD scale. By all means, this amounts to renouncing the idea that the mass function inferred by LIGO/Virgo events is primarily shaped by the physics of the early universe around the QCD phase transition. Nonetheless,  even in this case a significant bound (at the level $\fpbh\lesssim (0.75\div 2.4)\times 10^{-3}$ for $M_{\rm cut}=10^2\,M_\odot$, depending on the details of the mass function) is set by the non observations of mergers with a BH whose mass is sub-solar (see~\cite{Nitz:2022ltl,subsolar} and refs therein). In Fig.~\ref{fig5}, we recast this bound in terms of $f_{\rm GW}$ introduced in Eq.~\eqref{eq12}, obtaining  $f_{\rm GW}\lesssim 10^{-5}$. This illustrates how in the QCD-inspired scenario, PBHs can have at most a tiny contribution to the events detected by LIGO/Virgo, well below the  $10^{-3}$ level required in phenomenological fits~\cite{Franciolini:2021tla}. Note that the ``symmetric'' argument, i.e. the lack of mergers involving ``too heavy'' PBH, also puts constraints similar in strength to the CMB ones, but independent from them~\cite{Hutsi:2020sol}.

\section{Discussion and conclusions}\label{sectionIV}
In this paper we have assessed the viability of the scenario where a sizable primordial black hole (PBH) contribution to the LIGO/Virgo binary black hole mergers relies on a PBH mass spectrum  shaped by the QCD phase transition and related early universe evolution, down to the  $e^{+}e^{-}$ annihilation epoch. This scenario naturally enhances the probability to form PBH around the stellar mass scale, leading to PBHs up to hundreds of solar masses and, possibly, to the seeds of present-day supermassive black holes. 
The scenario is subject to numerous constraints, from peculiar CMB anisotropies induced by accretion onto PBHs to null results of gravitational wave searches for sub-solar BH mergers; from counts of supermassive black holes at high redshift to CMB spectral distortions. 
Our findings suggest that this scenario is not viable, unless, in practice, an ad hoc mass evolution for the PBH mass function and a a cutoff in power-spectrum very close to the QCD scale are introduced by hand. This obviously spoils the  ``naturaleness'' appeal of this scenario. 

Are there loopholes to the above conclusions?

As we mentioned, if the underlying PS contains further structures, instead of being flat as assumed in Eq.~\eqref{eq14}, the mass function would be altered. However, while quantitative differences at masses $M\gg 1\,M_\odot$ are to be expected (for an illustration, compare Fig.~\ref{fig3} based on Eq.~\eqref{eq13} with with Fig.~\ref{fig5} based on Eq.~\eqref{eq14}), this is not the case in the LIGO-Virgo mass range, due to the closeness to the QCD scale (unless, again, a sharp feature close to $k_{\rm QCD}$ is superimposed by hand).

Another assumption implicitly made is that the PBH mass function has not undergone a significant evolution, at least in its bulk properties.  Some mass function evolution is expected to happen as a consequence of accretion and mergers, which we discuss in turn. 

Unfortunately, existing attempts to take the effect of mass evolution via accretion into account, such as~\cite{DeLuca:2020fpg}, are not quantitatively reliable, since based on extrapolation of spherical, steady-state, cosmologically relevant solutions~\cite{Mack:2006gz,Ali-Haimoud:2016mbv} down to $z\lesssim 10$. Density, velocities, and accretion properties of PBH in the dark ages are instead expected to be dominated by the conditions in the assembling proto-halos, in turn requiring to assess baryonic structure formation in the non-linear regime for BH cosmologies in order to draw quantitative conclusions. However, let us entertain the possibility that the bulk of LIGO/Virgo merger events at tens of solar masses are actually the result of PBH seeds at $\sim 1\,M_\odot$ growing by more than one order of magnitude via accretion. We would be forced to conclude that the actual mass function inferred from LIGO/Virgo events would be rather determined by the unknown astrophysical history of accretion than by the cosmological initial mass function. Given current ignorance on dark ages and the reionisation period, one may then equally well invoke ``ordinary'' BH of $\sim 3-10\,M_\odot$ due to stellar collapses of first stars at $10\lesssim z\lesssim 20$, and similarly grown via accretion, as responsible for BH merger events involving masses of $\sim 60-80\,M_\odot$~\cite{LIGOScientific:2020iuh}. 
Perhaps more worrisome, this scenario would imply that a fraction of $10^{-3}$ of the total matter of the universe, i.e. close to $0.5\%$ of the total baryonic matter of the universe, would have been involved in accretion phenomena in the dark ages. This is a very large amount of material, roughly amounting to 10\% of the whole stellar production according to the inventory of ~\cite{Fukugita:2004ee}. It would appear rather astonishing that such a huge amount of matter would have left no observable electromagnetic trace, despite being heated up to high temperature in the accretion phenomenon. A more quantitative assessment of the viability of such a scenario awaits of course the elaboration of a concrete model.  

\begin{figure}[htbp!]
    \centering
    \includegraphics[width=0.45\textwidth]{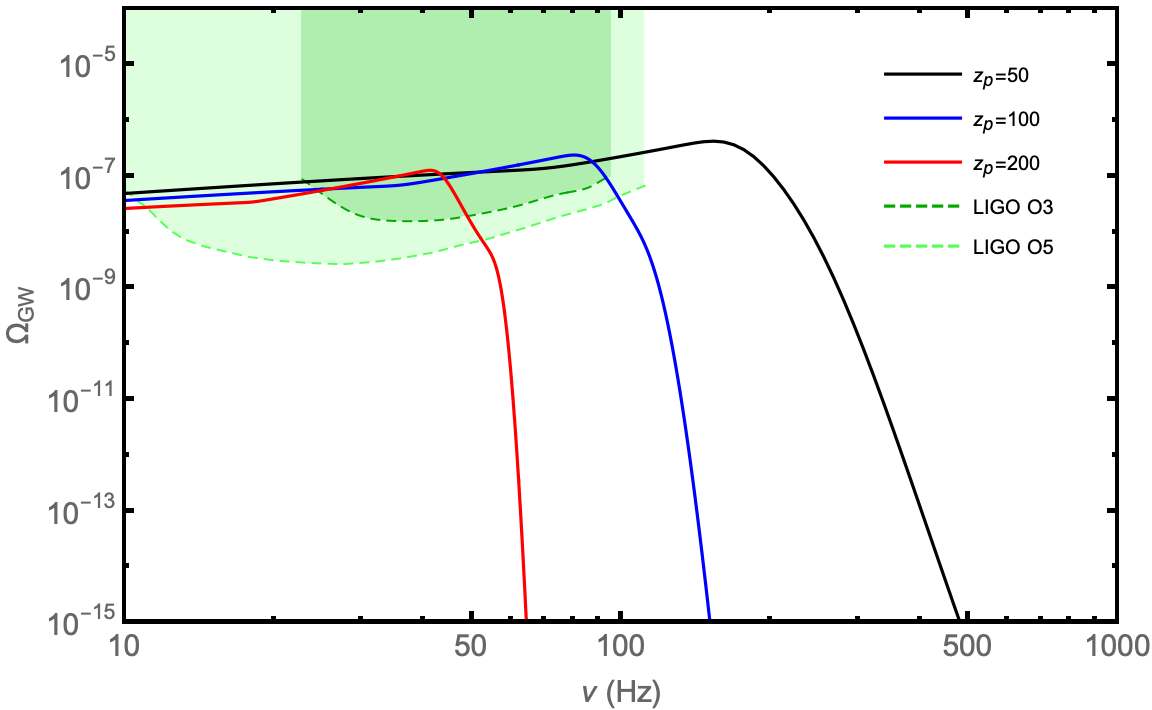}
    \caption{Stochastic gravitational wave background in the scenario where all solar PBHs merge once for different values of $z_{p}$ and $\sigma=5$, for $\fpbh=10^{-3}$. In dark green we also show the upper limits from the null searches for an isotropic SGWB from Advanced LIGO/Virgo O3~\cite{KAGRA:2021kbb}, while the lighter green shade represents the reach for run O5.}
    \label{fig6}
\end{figure}

The alternative possibility of major alterations due to mergers can be put on more solid quantitative grounds, at least at phenomenological level. First, it is rather difficult to build a model where the bulk of the PBH population undergoes one or more mergers (for concrete examples of the expected rates, typically detectable only with future generation detectors, see e.g.~\cite{Mukherjee:2021ags,Mukherjee:2021itf,Bagui:2021dqi}). Even ignoring that difficulty, however, such  a putative model would appear in conflict with observations.  In this scenario the amount of DM in the form of PBHs stays roughly constant, apart for a few percent of the mass converted in GWs at each merger. On average,  $\log_2(30)\simeq 5$ mergers would be needed in order for PBHs to shift their mass function from a peak at $\sim$1$\,M_\odot$  to a peak at $\sim$30$\,M_\odot$, thus explaining an ``anomalously heavy'' component of the LIGO/Virgo merger mass distribution. Yet, even a {\it single} merger on average for the whole population of PBH would lead to a tension with null searches of a stochastic gravitational wave background (SGWB) in LIGO/Virgo.  
We illustrate this point by computing  the SGWB following the formalism developed in~\cite{phinney2001practical}. The frequency spectrum writes
\begin{equation}
    \Omega_{\rm GW}(\nu)=\frac{\nu}{\rho_{c}}\int_{0}^{\nu_{\rm cut}}  \frac{N(z)}{1+z} \left( f_{r}\frac{{\rm d}E_{\rm GW}}{{\rm d}f_{r}} \right)  {\rm d}z,
    \label{eq15}
\end{equation}
where $f_{r}=f(1+z)$ is the redshift at the emission in terms of the one at the  the Earth $f$, $N(z)$ is number of events per comoving volume, $E_{\rm GW}$ is the energy emission in the form of GWs, and $\nu_{\rm cut}$ is a cutoff frequency. We compute these quantities for the merger of two PBHs of $1\,M_\odot$ using the formulae reported e.g. in section 4.6.7 of~\cite{franciolini2021primordial} (see also the phenomenological expression given in~\cite{PhysRevLett.106.241101}).
The result only mildly depends on the redshift dependence of the unknown merger rate $N(z)$. For illustration, let us consider a simple toy model characterized by a broadened peak at some redshift $z_{p}$, varied during the dark ages, parameterized via a Gaussian:
\begin{equation}
    N_{\rm bump}(z)=\frac{\fpbh \Omega_{\rm DM} \rho_{c}}{M_{\rm PBH} \sqrt{2 \pi \sigma^{2}}} \text{exp} \left[ -\frac{(z-z_{p})^{2}}{2 \sigma^{2}} \right],
    \label{eq16}
\end{equation}
with a finite width in redshift $\sigma=5$. The normalisation just imposes that a fraction $\fpbh$ of the DM mass has undergone mergers. In Figure~\ref{fig6}, we show the results for the benchmark value $\fpbh=10^{-3}$ and three values of $z_p$, compared with current bounds from~\cite{KAGRA:2021kbb} (darker shaded region) and projected sensitivity from the O5 run (lighter shaded region). Despite the roughness of the estimate and the unspecified details of the scenario, taking into account a normalisation about a factor 5 higher due to the multiple mergers, and an even more extended frequency span due to the broader mass function, clearly reveals that such a scenario is untenable.

In conclusion, the most appealing scenario to explain the required mass function to significantly contribute to LIGO/Virgo merger events, invoking the physics of the early universe between the QCD phase transition and the $e^\pm$ annihilation era~\cite{JEDAMZIK1998155,PhysRevD.59.124014,10.1093/mnras/stw2138,2018,Carr:2019hud,Carr:2019kxo,Jedamzik:2020omx}, does not appear viable. The possibility that some alternative mechanism is responsible for producing PBHs of stellar mass scale, with a much narrower mass function peaking between $10-100\,M_\odot$ is not ruled out if carefully designed to avoid CMB limits, but unfortunately way less predictive than the one analysed in this article. On the other hand, a marking feature of the proposed cosmological mechanism is its very wide PBH mass function, extending from 0.1-10 $M_\odot$ associated to the QCD phase transition, up to the $\sim 10^7\,M_\odot$ scale of SMBHs associated to the $e^\pm$ annihilation. Even if the contribution of these PBHs to the heavy tail of the LIGO/Virgo events were negligibile, it is still possible that (the seeds of) SMBHs inferred at high-$z$ are produced cosmologically. An example of such a scenario is provided by the dashed-gray curve in Fig.~\ref{fig4}. 
Its viability would still require that the tight nominal bounds from CMB spectral distortions can be relaxed to no more than $\fpbh(M\geq 10^6\,M_\odot)\sim 10^{-9}$ due e.g. to non-Gaussianities. In this case, it will be crucial to push the search for coalescing binaries involving possibly sub-solar mass BHs in future runs of the LIGO/Virgo/KAGRA facilities as well as at forthcoming GW detectors: These exotic events may thus provide a serendipitous smoking gun to understand the still mysterious origin of the heaviest BH in the universe, harbored at the core of most Galaxies. A further signal of (or constraint to) this scenario may come from the stochastic background in the nHz range probed by pulsar timing arrays, due to second order tensor perturbations generated by the scalar perturbations which produce the PBHs, similar to what discussed e.g. in~\cite{Inomata:2020xad}.

\begin{acknowledgments}
We would like to thank H. Veerm\"ae for comments, and Marco Taoso and Jens Chluba for feedback on the topics reported in Appendix~\ref{appA}. We are also grateful to Th{\'e}o Simon and Vivian Poulin for discussions on the topics reported in Appendix~\ref{appB}. 
\end{acknowledgments}
\bibliography{biblio}
\appendix

\newpage
\section{Considerations on CMB {\it spectral} distortions}\label{appA}
Since avoiding CMB anisotropy bounds on ${\cal O}(10-100)\, M_\odot$ PBH  requires a sufficiently suppressed power spectrum, the same mechanism would automatically get rid of CMB spectral distortions bounds applying to larger masses, roughly above 10$^{4}\,M_\odot$. Hence, in the main text we have not discussed bounds from CMB spectral distortions in detail. Here we revisit these bounds, commenting on the efficiency of another way sometimes discussed in the literature to soften the constraints, namely invoking non-gaussian pertubations~\cite{Nakama:2016kfq,Nakama:2017xvq}.

\begin{figure}[htbp!]
    \centering
    \includegraphics[width=0.45\textwidth]{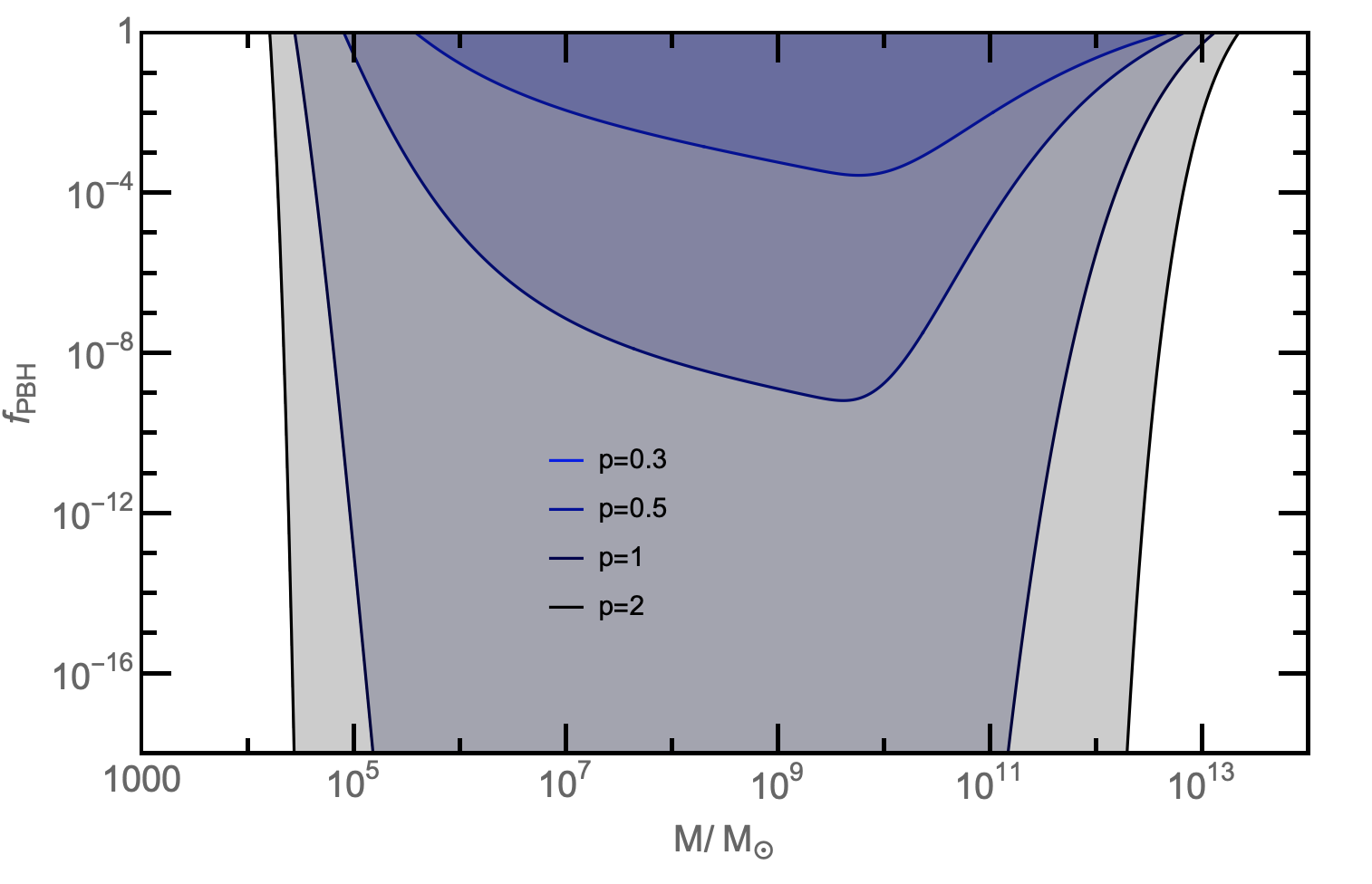}
    \caption{Excluded regions (shaded) of $\fpbh$ for a Dirac delta power spectrum (Eq.~\eqref{psdelta}) and for $p=\{0.3,0.5,1,2\}$.}
    \label{mubounds}
\end{figure}

\begin{figure}[htbp!]
    \centering
    \includegraphics[width=0.45\textwidth]{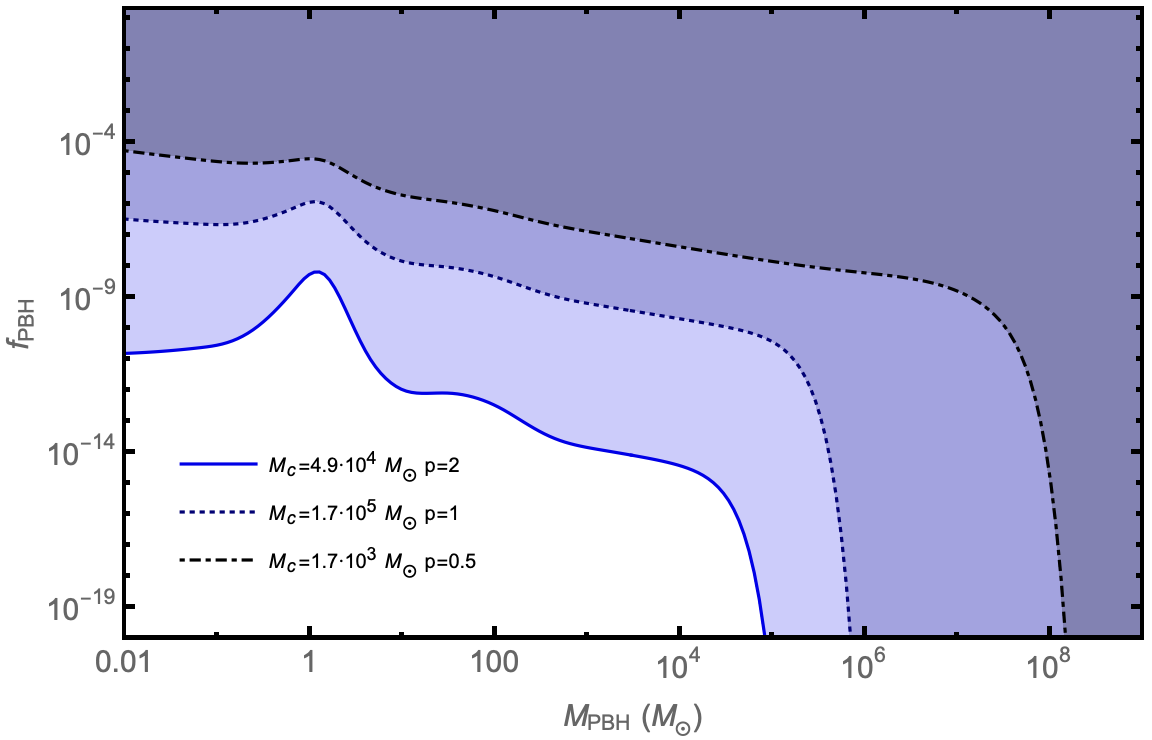}
    \caption{Excluded regions (shaded) of $\fpbh$ for the power spectrum in Eq.~\eqref{eq12}. Results are reported for three different values of the cutoff mass and degree of NG.}
    \label{figureextra}
\end{figure}

For the sake of definiteness, let us take the phenomenological probability distribution function of the primordial curvature perturbation $\zeta$ proposed in~\cite{Nakama:2016kfq}:

\begin{equation}
    P(\zeta)=\frac{1}{2\sqrt{2}\tilde{\sigma}\Gamma(1+1/p)} \text{exp} \left[ -\left( \frac{\mid \zeta \mid}{\sqrt{2}\tilde{\sigma}} \right)^{p} \right],
    \label{pdfzeta}
\end{equation}

where $\Gamma(x)$ is the gamma function and $p$ parametrizes the amount of NG, $p=2$ being the Gaussian case.\\
The variance  and the fraction of the Universe collapsing into PBH write respectively as

\begin{equation}
    \sigma^{2}=\int_{-\infty}^{\infty} \zeta^{2} P(\zeta) d\zeta=\frac{2 \Gamma(1+3/p)}{3 \Gamma(1+1/p)} \tilde{\sigma}^{2}
    \label{sigmas}
\end{equation}

\begin{equation}
    \beta=2 \int_{\zeta_{c}}^{\infty} P(\zeta)d\zeta=\frac{\Gamma(1/p,2^{-p/2} (\zeta_{c}/\tilde{\sigma})^{p})}{p\Gamma(1+1/p)},
    \label{beta}
\end{equation}

where $\zeta_{c}$ is the threshold for PBH formation, $\Gamma(x,a)$ is the incomplete gamma function and $\sigma=\tilde{\sigma}$ and $\beta=\text{erfc}(2^{-1/2}\zeta_{c}/\sigma)$ when $p=2$, which corresponds to the expected expression in the Gaussian case. Finally, in order to compute $\fpbh$, we use

\begin{equation}
    \fpbh=\left( \frac{M}{M_{\rm eq}} \right)^{-1/2} \frac{\beta(M)}{\Omega_{\rm DM}},
    \label{fpbh}
\end{equation}

where $M_{eq}=2.8\times 10^{17} M_{\odot}$.\\
\\
One can compute the $\mu$-distortion for a particular choice of the primordial curvature perturbation power spectrum $\mathcal{P}_{\zeta}(k)$ according to (see~\cite{Chluba:2012we})

\begin{eqnarray}
    \mu&=&2.2\int_{k_{min}}^{\infty} {\rm d}\ln k\, \mathcal{P}_{\zeta}(k)\times\nonumber \\ 
    &&\times \left[ \text{exp}\left(-\frac{k}{5400}\right)-\text{exp}\left(-\left[\frac{k}{31.6}\right]^{2}\right) \right]\, ,
    \label{mups}
\end{eqnarray}
where $k$ are measured in inverse Mpc.\\
Considering for simplicity a delta function power spectrum in $k$-space,
\begin{equation}
    \mathcal{P}_{\zeta}(k)=\sigma^{2}k\delta(k-k^{*}),
    \label{psdelta}
\end{equation}
the resulting $\mu$-distortion writes
\begin{equation}
    \mu=2.2 \sigma^{2} \left[ \text{exp}\left(-\frac{k^{*}}{5400}\right)-\text{exp}\left(-\left[\frac{k^{*}}{31.6}\right]^{2}\right) \right],
    \label{mudelta}
\end{equation}
where again $k$ is given in units of $\text{Mpc}^{-1}$.\\
Starting from the upper limit on the $\mu$-distortion set by FIRAS~\cite{Fixsen:1996nj}, that is $\mu \leq 9\times 10^{-5}$, 
one can set an upper bound on $\sigma^{2}$ via Eq.~\eqref{mudelta}. Then, one can obtain the corresponding value for $\tilde{\sigma}^{2}$ from Eq. \eqref{sigmas} and for a given value for the NG parameter $p$. Finally, plugging the result into Eq.~\eqref{beta} and, ultimately, Eq.~\eqref{fpbh}, one infers an upper bound on $\fpbh$. In Fig.s \ref{mubounds} and \ref{figureextra} we show the results for some scenarios with a different degree of NG, for the PS in Eq.~\eqref{psdelta} and Eq.~\eqref{eq12} respectively.

These results, which are equivalent to the ones reported in Figure 1 in~\cite{Nakama:2017xvq}, are specific to the particular choice of power spectrum in Eq.~\eqref{psdelta}. However, one could obtain equivalent results by taking any desired expression for the PS and applying Eq.~\eqref{mups}. In any case, Figure \ref{mubounds} illustrates the statement that large NGs soften the bound coming from $\mu$-distortions. 

One may wonder if one could soften the bounds arbitrarily strongly if going to more and more NG distributions. In what follows, we argue that this is not the case, since at some point one should take into account that the distorsions themselves are affected by the NG distribution. 

To illustrate this point, which to the best of our knowledge has been overlooked until now, let us review where Eq.~\eqref{mups} comes from. Following \cite{Chluba:2012we}, one can compute the $\mu$-distortion by applying Eq.~\eqref{musilk}:

\begin{equation}
    \mu=1.4 \frac{\Delta \rho_{\gamma}}{\rho_{\gamma}}=1.4 \int_{5\times 10^{4}}^{\infty} \mathcal{J}_{bb}(z) \frac{1}{\rho_{\gamma}}\frac{{\rm d}Q_{ac}}{{\rm d} z}{\rm d} z
    \label{musilk},
\end{equation}

which basically accounts for the fraction of the energy release ($Q_{ac}$) useful in causing a $\mu$-distortion. $\mathcal{J}_{bb}(z)$ is the visibility function for spectral distortions, reported for instance in~\cite{Chluba:2012gq}.

In order to compute the $\mu$-distortion one should compute the energy release rate and then apply Eq.~\eqref{musilk}. This computation, performed e.g. in~\cite{Chluba:2012gq} turns out to be non-trivial and rather technical, but an energy argument outlined in~\cite{Chluba:2012gq} yields essentially the correct result without passing through the Boltzmann equation. The reasoning is best illustrated by considering first the distortion associated to the mixing of two blackbodies at two slightly different temperatures, $T_{1}=T+\delta T\equiv T(1+\theta)$ and $T_{2}=T-\delta T\equiv T(1-\theta)$.
A 50\%-50\%  mixture of the two leads to a photon gas with average energy and number density given by

\begin{equation}
    \rho_{av}=a_{R}\frac{(T_{1}^{4}+T_{2}^{4})}{2}=a_{R} T^{4} \left[ 1+6\theta^{2}+\theta^{4} \right]\,
    \label{averagerho}
\end{equation}
\begin{equation}
    N_{av}=b_{R}\frac{(T_{1}^{3}+T_{2}^{3})}{2}=b_{R} T^{3} \left[ 1+3\theta^{2}\right],
    \label{averageN}
\end{equation}

where $a_{R}$ and $b_{R}$ are the well-known radiation constants (we follow the same notation as in~\cite{Chluba:2012gq}).\\
\\
Such an ensemble of photons clearly cannot be described as a blackbody: Its energy content $\rho_{av}$ will be higher than the energy content  $\rho_{BB}=a_{R} T_{BB}^{4}$ of the blackbody having the same number of photons, i.e. the one associated to the temperature $T_{BB}$ given by $N_{av}=b_{R}T_{BB}^{3}$.
The extra energy $\Delta \rho=\rho_{av}-\rho_{BB}$, given by
\begin{equation}
    \Delta \rho=a_{R}T^{4}\left( \left[ 1+6\theta^{2}+\theta^{4} \right]-\left[ 1+3\theta^{2}\right]^{4/3} \right),
\end{equation}
 is the one that can lead to spectral distortions via Eq.~\eqref{musilk}.
 
Generalising this argument, we expect that for small perturbations one has:

\begin{equation}
    \frac{\Delta \rho_{\gamma}}{\rho_{\gamma}} \simeq \frac{\rho_{av}-\rho_{BB}}{\rho_{pl}} \approx 2 \langle \theta^{2} \rangle  \implies \mu\simeq 2.8\langle \theta^{2} \rangle,
    \label{argresult}
\end{equation}
where in the last step we used Eq.~\eqref{musilk}. In the Gaussian case, this conclusion matches the result of a perturbative treatment of the Boltzmamnn equation at second order, as detailed in~\cite{Chluba:2012gq}.

However, in the context of PBH formation with sizable NG:
\begin{itemize}
    \item we know that PBH are born when a large overdensity collapses. Therefore, in this scenario, we can already foresee that PBH formation is associated to  temperature perturbations $\theta\gtrsim 0.1$, so one may question the validity of the aforementioned bound derived under the assumption that $\theta \ll 1$.
    \item if NGs are important, the results $\langle \theta^{2n+1} \rangle=0 $, $\langle \theta^{2n} \rangle\sim \left(\langle \theta^2\rangle\right)^n$ for integer $n$ that are valid for Gaussian variables may not hold. It is conceivable that a distribution has a comparatively smaller variance than the square root of its fourth moment, for instance. 
\end{itemize}
In order to assess this idea in a more quantitative way, we apply again the same energy argument discussed in the previous section but dropping the assumption $\theta \ll 1$ and keeping all the orders in $\theta$. This leads to a $\mu$-distortion of the form

\begin{eqnarray}
    \mu&=&1.4 \left( \left[ 1+6\langle\theta^{2}\rangle+4\langle\theta^{3}\rangle+\langle\theta^{4}\rangle \right]-\right.\nonumber\\
    &&\left.\left[ 1+3\langle\theta^{2}\rangle+\langle\theta^{3}\rangle\right]^{4/3} \right).
    \label{mucomplete}
\end{eqnarray}

Although this is not an exact computation of the $\mu$-distortion, we expect our results to hold up to $\mathcal{O}(1)$ factors.  
Note that the results recently obtained in~\cite{Acharya:2021zhq} suggest that the standard ``small distortion'' calculation (visibility function, etc.) remains valid up to sizable distorsions, and in the worst cases the approximation underestimates the actual effect.\\
From Eq.~\eqref{mucomplete} we can compute the $\mu$-distortion, provided we can compute moments of the temperature perturbation. The probability density function for such perturbations for the case at hand is however  linked to the curvature perturbation  via a non-linear relation. Without entering the details of this complicated subject, there are prescriptions on how the simplified formalism previously outlined should be generalized to take into account the critical nature of the collapse,  the finite size effect of the collapsing region leading to PBH, and the non-linearity of its relation with the fluctuation in the radiation density. For instance, Appendix C of~\cite{Wu:2021mwy}, provides a compact review.

If $F$ denotes the variable associated to the curvature perturbation~\footnote{In practice, $F\equiv 4\tilde{ \zeta} \equiv 4 r_m\partial_r \zeta(r_m)$ and $r_m$ is the local maximum of the so-called {\it compaction function}, which is what truly determines the PBH collapse in such a more advanced treatment.} that is actually distributed according to Eq.~\eqref{pdfzeta}, the corresponding pdf for $\theta$ accounting for the non-linearity can be obtained via the following relations:

\begin{equation}
    \frac{\delta \rho}{\rho}  =F-\frac{3}{8}F^{2} \quad \text{and}    \quad \theta  =\frac{1}{4}\frac{\delta \rho}{\rho},
\end{equation}
 \\
The pdf for $\theta$, then, becomes
\begin{eqnarray}
    P(\theta)&=&\frac{1}{\sqrt{1-6 \theta}}\frac{1}{2\sqrt{2}\tilde{\sigma}\Gamma(1+1/p)}\times\nonumber\\ 
    &&\exp \left[ -\left( \frac{\mid \frac{1}{3} (1\pm \sqrt{1-6\theta}) \mid}{\sqrt{2}\tilde{\sigma}} \right)^{p} \right],
    \label{pdftheta}
\end{eqnarray}
where the +/- solution is to be integrated over the range $\{-\infty,1/6\}$. Note that once we fix the amount of NG (that is, the parameter $p$) the only parameter left is $\tilde{\sigma}$, the typical width of the distribution (for $\tilde \zeta$), which will enter into the computation of $\fpbh$.\\

\begin{figure}[!th]
    \centering
    \includegraphics[width=0.45\textwidth]{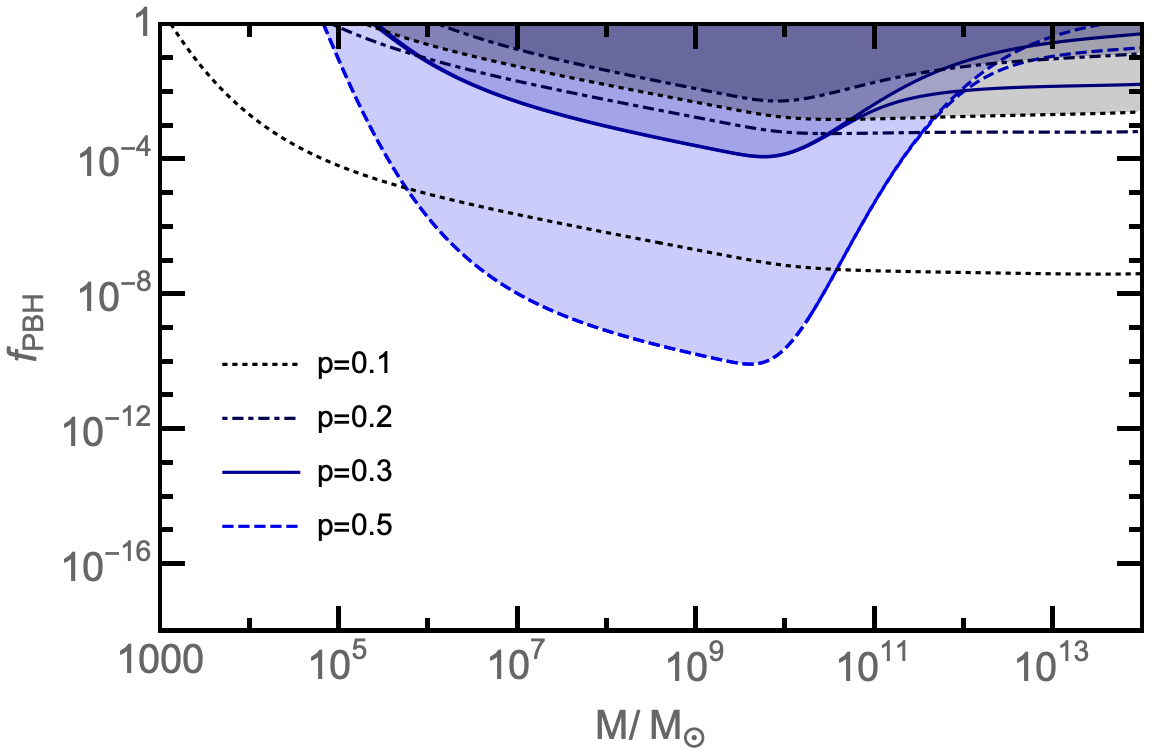}
    \caption{Upper bounds on $\fpbh$ for $p=\{0.1,0.2,0.3,0.5\}$. The shaded areas correspond to the already existing results obtained by neglecting higher orders in $\theta$, while the bounds without filling correspond to the new results obtained from Eq.~\ref{mucomplete}.}
    \label{munew}
\end{figure}

We can now find which is the value of $\tilde{\sigma}$ that saturates the FIRAS bound for a particular choice of $p$ and use it to compute the corresponding upper limit on $\fpbh$. We show the results in Figure \ref{munew}, where we apply the same  exponential cutoff function as in Eq.~\eqref{mudelta}, in order to compare the results with those in Fig.~\ref{mubounds}. A more correct calculation according to Eq.~\eqref{mups} is prevented by the fact that from the parametric toy model of the pdf in Eq.~\eqref{pdfzeta} one cannot infer
the PS (technically, the pdf depends on all cumulants, as  illustrated via the Gram–Charlier or Edgeworth series expansion).

For values  $p \gtrsim 0.3$, the bounds displayed in Figure~\ref{mubounds} are almost unmodified. However, this is not the case for lower values of $p$. Indeed, as we can observe in Figure \ref{munew}, there is a significant enhancement due to higher order terms of $\sim 1$ and $\sim 5$ orders of magnitude in the case of $p=0.2$ and $p=0.1$, respectively.
Although our considerations are not rigorous and based on an energy argument, they illustrate the point that carefulness should be applied when invoking too large NGs to dismiss spectral bounds from CMB, since eventually NG corrections to the CMB spectral bound may more than compensate the softening of the bound naively associated to the NGs.

\section{CMB bounds on disk-accreting PBH including extended mass functions }\label{appB}

In this Appendix, we derive CMB constraints on disk-accreting PBH following the analysis of \cite{accretion} and \cite{Serpico:2020ehh}, but including for the first time the presence of very extended mass functions (as the one considered in this work). The main goal is to compare the resulting bounds with those obtained by the approximate method that was outlined in Section~\ref{sectionIIC}. \ 

Accretion of matter onto massive PBH leads to the emission of high-energy radiation, capable of altering the thermal and ionization history of the universe. Consequently, this affects the shape of the CMB temperature and polarization anisotropy spectra. This effect is parameterized through the energy injection rate per unit volume. For a certain PBH mass $M$ and redshift $z$, it is given by:
\begin{equation}
\frac{{\rm d}E}{{\rm d}V{\rm d}t}(z, M)\Bigg|_{\rm inj}  = L_{\rm{acc}}(z,M) f_{\rm PBH} \frac{\rho_{\rm DM} (z)}{M},
\label{energy_inj}
\end{equation}
where $L_{\rm{acc}}$ is the accretion luminosity, which is tightly correlated with the assumed accretion geometry (see \cite{Ali-Haimoud:2016mbv} and \cite{accretion} for the calculation of $L_{\rm{acc}}$ assuming spherical- and disk-like accretion, respectively). To compute the impact on the CMB, one still needs to describe what amount of the injected energy is deposited in the surrounding medium, either through heating, ionization or excitation of the hydrogen atoms. This information is encoded in the energy deposition function per channel $f_c (z, M)$. In addition, one has to add the contributions to the deposited energy from all PBH masses, which are weighted by the mass function $\psi_p (M)$\footnote{Throughout this appendix, we consider a slightly different convention for the extended mass function $\psi_p(M)$. In particular, we fix the amplitude $\Delta =10^{-2}$ and add a normalization to $\psi_p$ such that $\int dM \psi_p(M)=1$. We do this because, for each value of the cutoff mass $M_{\rm cut},$ we will be putting constraints directly on $f_{\rm PBH}$, rather than on $\Delta$. We also assume the horizon entry prescription for computing $\delta_c$, instead of the time average one.}. Hence, the total energy deposition rate per unit volume per channel is:
\begin{equation}
\frac{{\rm d}E}{{\rm d}V{\rm d}t}(z)\Bigg|_{\rm dep, c}  = \int dM f_c (z, M) \psi_p (M) \frac{{\rm d}E}{{\rm d}V{\rm d}t}\Bigg|_{\rm inj} ,
 \label{energy_dep}
\end{equation}
where the integral is to be performed over all masses relevant for the extended mass function under consideration. 
For a monochromatic mass function $\psi_p (M) = \delta (M- M_c)$, the integral in  Eq.~\eqref{energy_dep} is performed trivially, yielding to the expressions shown in Eqs. (19, 20) of \cite{accretion}.  In the case of an extended mass function as the one shown in  Eq.~\eqref{eq8}, the integral has to be performed numerically. \ 

\begin{figure}[!th]
    \centering
    \includegraphics[width=0.5\textwidth]{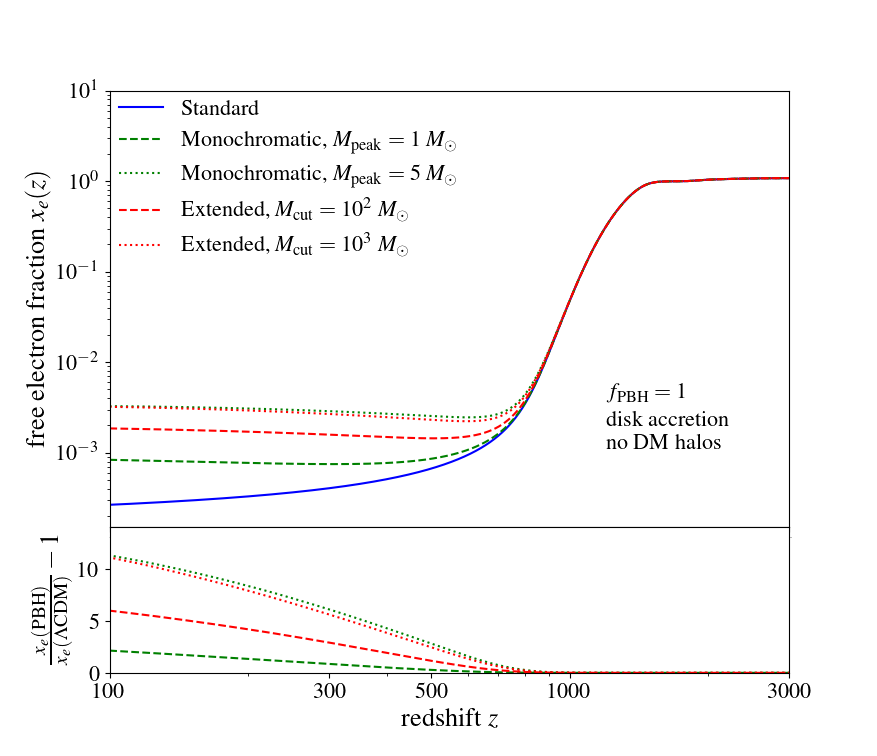}
    \caption{Comparison of the free electron fraction for different configurations of the accreting PBH mass functions. We take $\fpbh=1$ and always assume disk-like accretion without including the formation of DM halos. The curve labelled ``Standard'' refers to the $\Lambda$CDM model with parameters set to the best-fit of \emph{Planck} 2018 TTTEEE+lowE+lensing~\cite{planck2020}.   }
    \label{xe_PBH}
\end{figure}

\begin{figure}[!th]
    \centering
    \includegraphics[width=0.5\textwidth]{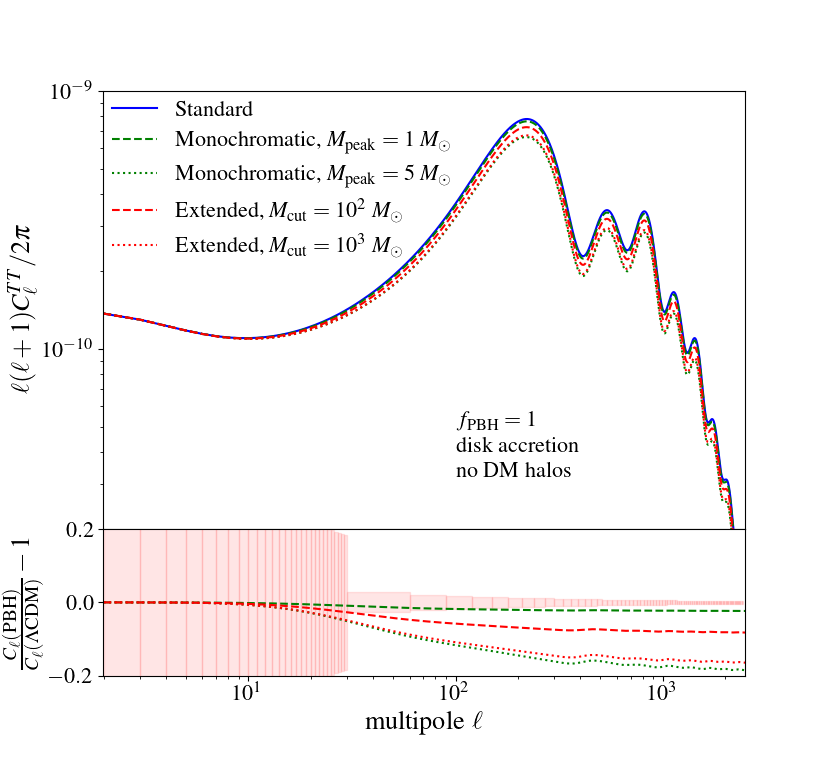}
    \caption{Same as in Figure \ref{xe_PBH}, but for the CMB TT power spectrum.}
    \label{clTT_PBH}
\end{figure}

We have implemented Eq.~\eqref{energy_dep} in our modified version of the branch \texttt{ExoCLASS} \cite{Stocker:2018avm} of the public Boltzmann solver \texttt{CLASS} \cite{Blas:2011rf}. Our code is publicly available at \url{https://github.com/GuillermoFrancoAbellan/ExoCLASS_extPBH}. The code gets the information about $\psi_p (M)$ by reading a pre-computed table for different values of the cutoff mass $M_{\rm cut}$ and for a fixed cosmology: $\Omega_{\rm cdm}=0.2645$, $A_{s}=2.0989\times 10^{-9}$, $n_{s}=0.9649$. These values correspond to the $\Lambda$CDM best-fit values from the \emph{Planck} 2018 TTTEEE+lowE+lensing analysis~\cite{planck2020}. Fixing the cosmology for the calculation of $\psi_p(M)$ should not be a problem when doing Monte Carlo Markov Chain (MCMC) analysis, since the very small values of  $f_{\rm PBH}$ that are allowed by CMB data are not expected to produce significant shifts in the $\Lambda$CDM parameters. The implementation of the extended mass function is more computationally demanding than the monochromatic case, since a single \texttt{ExoCLASS} run now requires calling the python module \texttt{DarkAges} (in charge of computing $f_c (z, M)$) several times, one per each PBH mass.  For this reason, we choose to compute the integral in Eq.~\eqref{energy_dep} using $\sim 50$ mass bins, which provides the speed yet accurate enough for the purposes of the current analysis. \ 

In Figures \ref{xe_PBH} and \ref{clTT_PBH} we compare the free electron fraction and the CMB TT power spectrum for several configurations of the PBH mass function. We observe that the extended mass function generically produces stronger effects on $x_e(z)$ (and consequently on $C_\ell^{TT}$) than a monochromatic mass function located at $\sim 1 \ M_{\odot}$. Indeed, even if the extended mass function under consideration is peaked around that mass, it has also a support at higher masses,  for which the impact of accretion is quite large. This is also the reason why the effects of the extended mass function become stronger for larger values of $M_{\rm cut}$. Nevertheless, the extended mass function does not introduce any new signatures as compared to the monochromatic case, and we can actually find configurations for which the effects of the monochromatic and extended mass functions are almost identical (see red and green dotted curves in Figures \ref{xe_PBH} and \ref{clTT_PBH}) For this reason, we anticipate that our constraints wont be significantly different from those obtained with the simple recasting of the existing monochromatic bounds.

To derive the 95 \% CL bounds, we run a MCMC using the public code \texttt{MontePython-v3} \cite{Brinckmann:2018cvx} interfaced with our modified version of \texttt{ExoCLASS}. We perform the analysis with a Metropolis-Hasting algorithm, assuming flat priors on $\{\omega_b, \omega_{\rm cdm}, H_0, \ln(10^{10}A_s), n_s, z_{\rm reio}, f_{\rm PBH}\}$ at two different cutoff masses, $M_{\rm cut} / M_{\odot} = [10^2, 10^{4.5}]$. We adopt the \emph{Planck} collaboration convention in modelling free-streaming neutrinos as two massless species and one massive with $m_{\nu} = 0.06 \ \rm{eV}$. We include the same data sets as in \cite{Serpico:2020ehh}. Namely, we use data from the \emph{Planck} 2018 high-$\ell$ and low-$\ell$ TT, EE and lensing~\cite{planck2020}; the isotropic BAO measurements from 6dFGS at $z=0.106$ \cite{Beutler:2011hx} and from the MGS galaxy sample of SDSS at $z=0.15$ \cite{Ross:2014qpa}; the anisotropic BAO and the growth function $f\sigma_8(z)$ measurements from the CMASS and LOWZ galaxy samples of BOSS DR12 at $z=0.38, 0.51$ and $z=0.61$ \cite{BOSS:2016wmc}. In addition, we use the Pantheon supernovae dataset including measurements of the luminosity distances of 1048 SNe Ia in the redshift range $0.01 < z < 2.3$ \cite{Pan-STARRS1:2017jku}. Our runs assume disk accretion, but we neglect the presence of DM halos around the PBH for simplicity. Once we obtain our bounds on $f_{\rm PBH}$, we convert them into $f_{\rm GW}$ using Eq. \eqref{eq12}.\ 

\begin{table}[!th]
\addtolength{\tabcolsep}{+4pt}
\renewcommand{\arraystretch}{1.5}
\begin{tabular}{|l|l|c|c|}
\hline
\multicolumn{1}{|c}{} & \multicolumn{1}{c|}{} & $f_{\rm PBH}^{\rm max}$  & $f_{\rm GW}^{\rm max}$ \\
\hline
\multirow{2}{*}{$M_{\rm cut} =10^2 \ M_{\odot}$} & Full & $0.129$ & $2.83\times 10^{-3}$\\
\cline{2-4}
& Approx & $0.177$ & $3.88\times 10^{-3}$ \\
\hline
\multirow{2}{*}{$M_{\rm cut} =10^{4.5} \ M_{\odot}$} & Full & $1.99\times 10^{-3}$ & $4.87\times 10^{-5}$ \\
\cline{2-4}
& Approx & $ 3.09\times 10^{-3}$ & $7.54\times 10^{-5}$ \\
\hline
\end{tabular}
\caption{The 95\% C.L. limits on $f_{\rm PBH}$ and $f_{\rm GW}$ assuming disk accretion {\it and no DM halos}. These results are obtained for two different cutoff masses and two different methodologies. ``Full'' refers to the bounds that are obtained by a careful modelling the effects of the extended mass function on the CMB spectra, while ``Approx' refers to the quick recasting of the existing monochromatic bounds using Eq. \eqref{eq11}. }
\label{tab:f_pbh_comp}
\end{table}

In Table \ref{tab:f_pbh_comp} we compare the bounds on $f_{\rm PBH}$ and $f_{\rm GW}$ obtained with either the dedicated numerical calculation (to which we refer as ``Full'') or with the recasting of the monochromatic bounds (to which we refer as ``Approx''). We also compare the bounds for two different values of the cutoff mass, $M_{\rm cut} / M_{\odot} = [10^2, 10^{4.5}]$. Interestingly, we find that the bounds on $f_{\rm PBH}$ and $f_{\rm GW}$ from the ``Full'' method are a factor $\sim 1.4-1.5$ stronger than those from the ``Approx'' method, with the improvement in the bound growing slightly when more extended mass functions are considered. 
We conclude that the CMB constraints on extended PBH mass functions using the recasting of the monochromatic bounds, as done in the main text, leads to {\it conservative} bounds. Dedicated numerical calculations for extended mass function would however be needed once one could achieve a better understanding of the PBH accretion physics, which remains at the moment the dominant theoretical uncertainty.

\end{document}